# $N = 2$ Chiral Supergravity in $(10 + 2)$-Dimensions As Consistent Background for Super $(2 + 2)$-Brane [1]

Hitoshi NISHINO

*Department of Physics*
*University of Maryland*
*College Park, MD 20742-4111, USA*
E-Mail: nishino@umdhep.umd.edu

### Abstract

We present a theory of $N = 2$ chiral supergravity in $(10 + 2)$-dimensions. This formulation is similar to $N = 1$ supergravity presented recently using null-vectors in 12D. In order to see the consistency of this theory, we perform a simple dimensional reduction to ten-dimensions, reproducing the type IIB chiral supergravity. We also show that our supergravity can be consistent background for super $(2 + 2)$-brane theory, satisfying fermionic invariance of the total action. Such supergravity theory without manifest Lorentz invariance had been predicted by the recent F-theory in twelve-dimensions.

---

[1] This work is supported in part by NSF grant # PHY-93-41926.

## 1. Introduction

Recently we have constructed an $N=1$ supergravity theory in twelve-dimensions (12D) with the signature $(10,2)$ [1], motivated by the development of F-theory in 12D [2][3][4]. However, this $N=1$ theory was not yet a weak coupling (field theory) limit of F-theory with the maximal symmetry, due to the lack of the maximal $N=2$ supersymmetry. In fact, the supergravity theory corresponding to F-theory is supposed to reproduce type IIB supergravity in 10D [5], while $N=1$ supergravity in 12D [1] produces only the $N=1$ supergravity sector of type I or heterotic superstring in 10D. To put it differently, as the maximal $N=1$ supergravity in 11D corresponds to M-theory [6], the $N=2$ chiral supergravity in 12D corresponds to F-theory. The recent developments of super $(2+2)$-brane formulation in *flat* superspace [7] as well as S-theory [8], or higher-dimensional theories with two time coordinates [9], all suggest the existence of consistent formulation of Lorentz non-invariant $N=2$ supergravity in 12D, and it seems imperative to establish the field theory of $N=2$ chiral supergravity in 12D.

In this paper we establish a component formulation of $N=2$ chiral supergravity in 12D for the first time. This result is based on our previous experience with $N=1$ supergravity [1] as well as supersymmetric Yang-Mills theory [10] in 12D. Our $N=2$ supergravity shares many aspects with the previous $N=1$ supergravity in 12D [1], such as the field content similar to type IIB theory in 10D [5], with no additional tensor fields with ranks higher than those in 10D [1], the elaborate usage of null-vectors [1], a pair of gravitini each in the Majorana-Weyl representation [1], lack of invariant lagrangian [1], or the chirality structure in 12D parallel to that in 10D [5]. We will see how the self-duality condition for the fifth-rank antisymmetric field strength in 10D [5] is "oxidized" in 12D with peculiar involvement of a null-vector. After extracting lower-dimensional superspace constraints from the component result, we also show the fermionic $\kappa$ [11][12] and $\eta$-invariances [1] of the total action for super $(2+2)$-brane on our superspace background, under an extra constraint imposed on 12D supercoordinate variables. We see that the $\eta$-symmetry deletes unwanted degrees of freedom in the super $(2+2)$-brane, to accord with the type IIB theory in 10D. Our $N=2$ supergravity theory most likely provides far-reaching techniques for many applications, from the curved 12D superspace background for super $(2+2)$-brane theory [7], to analysis of compactifications of F-theory [13], with possible solution for cosmological constant problem in 4D *via* strong/weak duality between 4D and 3D [2][14].

## 2. Preliminaries

Before presenting our results, we prepare our notational conventions, in particular with our null-vectors. Most notations in this paper are the same as those in refs. [1] and [10].

First of all, we note that our signature is $(\eta_{mn}) = \text{diag.}(-,+,+,\cdots,+,+,-)$ [1][10], where $m, n, \ldots = (0), (1), \ldots, (9), (11), (12)$ are for local Lorentz coordinates, distinguished from the curved ones $\mu, \nu, \ldots = 0, 1, \ldots, 9, 11, 12$. Due to the $N=2$ feature of the system, we need both of the constant null-vectors defined by [1][10]

$$(n_m) = (0, 0, \cdots, 0, +\tfrac{1}{\sqrt{2}}, +\tfrac{1}{\sqrt{2}}) \ , \quad (m_m) = (0, 0, \cdots, 0, +\tfrac{1}{\sqrt{2}}, -\tfrac{1}{\sqrt{2}}) \ , \tag{2.1}$$

It is convenient to define $\pm$-indices for the extra dimensions for an arbitrary vector $V_m$ by

$$V_\pm \equiv \tfrac{1}{\sqrt{2}} \left( V_{(11)} \pm V_{(12)} \right) \ . \tag{2.2}$$

Accordingly, we have $n_+ = n^- = m_- = m^+ = +1$, $n_- = n^+ = m_+ = m^- = 0$, as in [1]. As in the $N=1$ case, we also need the combination of the null-vectors with the Dirac matrices [1], such as

$$\not{n} \equiv n_m \gamma^m \ , \quad \not{m} \equiv m_m \gamma^m \ , \quad \not{n}^2 = \not{m}^2 = 0 \ , \quad P_\uparrow \equiv \tfrac{1}{2} \not{n}\not{m} \ , \quad P_\downarrow \equiv \tfrac{1}{2} \not{m}\not{n} \ ,$$
$$P_\uparrow^2 = P_\uparrow \ , \quad P_\downarrow^2 = P_\downarrow \ , \quad P_\uparrow P_\downarrow = P_\downarrow P_\uparrow = 0 \ , \quad P_\uparrow + P_\downarrow \equiv I \ , \quad P_{\uparrow\downarrow} \equiv P_\uparrow - P_\downarrow \ . \tag{2.3}$$

In particular, $P_\uparrow$ and $P_\downarrow$ play roles of projection operators in the extra space in the representations of spinors in 12D, when performing dimensional reductions into 10D [1]. Also to be mentioned is the important feature of our Lorentz generators [1]

$$(\widetilde{\mathcal{M}}_{mn})^{rs} \equiv +2\widetilde{\delta}_{[m}{}^r \widetilde{\delta}_{n]}{}^s \quad (\widetilde{\delta}_-{}^m = 0, \text{ otherwise } \widetilde{\delta}_m{}^n = \delta_m{}^n) \ , \tag{2.4a}$$

$$(\widetilde{\mathcal{M}}_{mn})_\alpha{}^\beta \equiv +\tfrac{1}{2} (\gamma_{mn} P_\uparrow)_\alpha{}^\beta \ , \quad (\widetilde{\mathcal{M}}_{mn})_{\dot\alpha}{}^{\dot\beta} \equiv +\tfrac{1}{2} (P_\downarrow \gamma_{mn})_{\dot\alpha}{}^{\dot\beta} \ , \tag{2.4b}$$

where the first one has vanishing components for any index $m, n, r, s = -$, while the second line has extra $P_\uparrow$ and $P_\downarrow$. Note that $\alpha, \beta, \ldots$ (or $\dot\alpha, \dot\beta, \ldots$) are spinorial indices with negative (or positive) chirality.[2] Another important property is their action on spinorial indices, e.g.,

$$(\mathcal{M}_{mn} \circ \psi)_\alpha = (\mathcal{M}_{mn})_\alpha{}^\beta \psi_\beta = \tfrac{1}{2} (\gamma_{mn} P_\uparrow)_\alpha{}^\beta \psi_\beta \ ,$$
$$(\mathcal{M}_{mn} \circ \psi)^\beta = -\psi^\alpha (\mathcal{M}_{mn})_\alpha{}^\beta = -\tfrac{1}{2} \psi^\alpha (\gamma_{mn} P_\uparrow)_\alpha{}^\beta \ , \tag{2.5}$$

which was not explicitly given in [1]. Due to these peculiar features of Lorentz generators, all the superspace Bianchi identities are satisfied in $N=1$ supergravity [1], even though some superficially "constant" matrices such as $(\gamma^m)_{\alpha\beta}$ are not really invariant under these generators. This is the price we have to pay, for violating manifest Lorentz covariance in the total 12D [1], and this is the same in our present $N=2$ case. The analog of Bianchi identities in component formulation is the closure of gauge algebra, as will be discussed.

---

[2]Therefore in this paper our gravitino with negative chirality has *undotted* index $\psi_\mu{}^\alpha$, while our gravitello with positive chirality has *dotted* index $\lambda_{\dot\alpha}$, as will be mentioned. The *bars* we use in this paper denote the Dirac conjugate [15] as in (2.13), but not *dotted* spinors.

We next briefly outline the geometry for the scalar coset $SU(1,1)/U(1)$ as in [5]. The scalar fields $V_\pm{}^\alpha$ are $SU(1,1)$ group matrix-valued,[3] and transform as

$$\delta V_\pm{}^\alpha = m^\alpha{}_\beta V_\pm{}^\beta \pm i\Sigma V_\pm{}^\alpha \ , \tag{2.6}$$

under the infinitesimal global $SU(1,1)$ transformation with the constant parameter

$$(m^\alpha{}_\beta) = \begin{pmatrix} i\gamma & \alpha \\ \alpha^* & -i\gamma \end{pmatrix} \ , \tag{2.7}$$

and the local infinitesimal $U(1)$ transformation with the real parameter $\Sigma$. Since the $V$'s satisfy the relationships

$$\epsilon_{\alpha\beta} V_-{}^\alpha V_+{}^\beta = \det V = 1 \ , \quad V_-{}^\alpha V_+{}^\beta - V_+{}^\alpha V_-{}^\beta = \epsilon^{\alpha\beta} \ , \tag{2.8}$$

we do not need their inverse matrix. As usual in the coset formulation, we need the composite $U(1)$ connection

$$Q_\mu = -i\epsilon_{\alpha\beta} V_-{}^\alpha \partial_\mu V_+{}^\beta \tag{2.9}$$

transforming as $\delta Q_\mu = \partial_\mu \Sigma$, and the $SU(1,1)$ invariant field strength $P_\mu = -\epsilon_{\alpha\beta} V_+{}^\alpha \partial_\mu V_+{}^\beta$ transforming as $\delta P_\mu = 2i\Sigma P_\mu$. Relevantly, among the fields in our supergravity multiplet $(e_\mu{}^m, \psi_\mu, A_{\mu\nu\rho\sigma}, \lambda, A_{\mu\nu}{}^\alpha, V_\pm{}^\alpha)$, the following fields transform under $SU(1,1) \otimes U(1)$:

$$\delta A_{\mu\nu}{}^\alpha = m^\alpha{}_\beta A_{\mu\nu}{}^\beta \ , \quad \delta\psi_\mu = \tfrac{i}{2}\Sigma\psi_\mu \ , \quad \delta\lambda = \tfrac{3i}{2}\Sigma\lambda \ . \tag{2.10}$$

We have field strengths associated with the $V_\pm{}^\alpha$, which will be presented shortly.

There is a remark regarding our $SU(1,1)$ in our formulation. According to the original prediction of F-theory [2][16], the $SU(1,1)$ symmetry in type IIB is to be generated *e.g.*, upon toroidal compactifications from 12D into 10D. Our system, however, has this symmetry as a classical symmetry from the outset. At the present time, we do not have any alternative formulation in which this symmetry comes out upon compactification, and neither do we know if supergravity formulation itself is unique in 12D. This is still an unsolved problem yet to be investigated in the future.

We finally give the crucial notations/relations for the manipulation of spinors in 12D [15]. For the chirality we need $\gamma_{13} \equiv \gamma_{(0)}\gamma_{(1)}\cdots\gamma_{(9)}\gamma_{(11)}\gamma_{(12)}$, so that

$$\gamma^{[N]} = \frac{(-1)^{N(N-1)/2}}{(12-N)!} \epsilon^{[N][12-N]} \gamma_{[12-N]}\gamma_{13} \ . \tag{2.11}$$

Here the index $[N]$ is used for normalized totally antisymmetric indices, *e.g.*, $G^{[3]}G_{[3]} \equiv G^{\mu\nu\rho}G_{\mu\nu\rho}$, and the $\epsilon$-tensor is defined by $\epsilon^{01\cdots 9\,11\,12} = +1$. There are technically important relationships for Weyl spinors, such as

$$\gamma^{[6]}\psi_+ S_{[6]} \equiv 0 \ , \quad \gamma^{[6]}\psi_- A_{[6]} \equiv 0 \ , \tag{2.12}$$

---

[3]The indices $\alpha, \beta, \cdots = 1, 2$ are not to be confused with the 12D spinorial indices in (2.4) and (2.5), as long as they are clear from the context.

where $S_{[6]} \equiv +(1/6!)\,\epsilon_{[6]}{}^{[6]'} S_{[6]'}$ and $A_{[6]} \equiv -(1/6!)\,\epsilon_{[6]}{}^{[6]'} A_{[6]'}$ are self-dual and anti-self-dual tensor in 12D, while $\gamma_{13}\psi_\pm = \pm\psi_\pm$.

Other frequently used flipping property [15] for two Weyl spinors $\psi_1$ and $\psi_2$ is

$$\left(\overline{\psi}_1 \gamma^{\mu_1\cdots\mu_N} \psi_2\right) = +\left(\overline{\psi}_2 \gamma^{\mu_N\cdots\mu_1} \psi_1\right) = (-)^{N(N-1)/2}\left(\overline{\psi}_2 \gamma^{\mu_1\cdots\mu_N} \psi_1\right) \ . \tag{2.13}$$

Another important relations are such as the hermitian conjugation [15] for two Weyl spinors $\psi_1$ and $\psi_2$:

$$\begin{aligned}
\left(\overline{\psi}_1 \gamma^{\mu_1\cdots\mu_N} \psi_2\right)^\dagger &= \psi_2^\dagger \left(\gamma^{\mu_1\cdots\mu_N}\right)^\dagger \overline{\psi}_1{}^\dagger = +\left(\overline{\psi}_2 \gamma^{\mu_N\cdots\mu_1} \psi_1\right) \\
&= (-1)^{N(N-1)/2}\left(\overline{\psi}_2 \gamma^{\mu_1\cdots\mu_N} \psi_1\right) = +\left(\overline{\psi}_1^{\,*} \gamma^{\mu_1\cdots\mu_N} \psi_2^*\right) \ .
\end{aligned} \tag{2.14}$$

Here $*$-symbol is a complex conjugation of a Weyl spinor, i.e., $\psi^* = \left(\psi^{(1)} + i\psi^{(2)}\right)^* = \psi^{(1)} - i\psi^{(2)}$ for two Majorana-Weyl spinors $\psi^{(1)}$ and $\psi^{(2)}$ forming a Weyl spinor $\psi$.

### 3. $N = 2$ Supergravity in 12D

Our field content for our $N = 2$ supergravity is $(e_\mu{}^m, \psi_\mu, A_{\mu\nu\rho\sigma}, \lambda, A_{\mu\nu}{}^\alpha, V_\pm{}^\alpha)$, where as usual, $e_\mu{}^m$ is the zwölfbein, $\psi_\mu$ is a Weyl gravitino, i.e., a pair of Majorana-Weyl spinors in 12D with $\gamma_{13}\psi_\mu = -\psi_\mu$ [15], $A_{\mu\nu\rho\sigma}$ is the fourth-rank tensor subject to the anti-self-duality condition (3.7d), $A_{\mu\nu}{}^\alpha$ is a complex second-rank antisymmetric tensor with the curved coordinate indices on the coset $SU(1,1)/U(1)$ [5], $\lambda$ is a pair of Majorana-Weyl spinors satisfying $\gamma_{13}\lambda = +\lambda$ which we call "gravitello" for convenience. The $V_\pm{}^\alpha$ are the scalar fields parametrizing the coset $SU(1,1)/U(1)$ with the $U(1)$ charges $\pm$. As has been already clear, this field content is parallel to the 10D case, in particular, the scalars for the coset $SU(1,1)/U(1)$.

The supersymmetry transformation rule for our multiplet is

$$\delta e_\mu{}^m = \left[(\overline{\epsilon}\gamma^{mn}\psi_\mu)n_n + (\overline{\epsilon} P_{\uparrow\downarrow}\psi_\mu)n^m\right] + \text{c.c.} \ , \tag{3.1a}$$

$$\delta\psi_\mu = \widehat{D}_\mu\epsilon - \frac{i}{480}(P_\downarrow\gamma^{[5]}\gamma_\mu\epsilon)\widehat{F}_{[5]} + \frac{1}{96} P_\downarrow\left(\gamma_\mu{}^{[3]}\widehat{G}_{[3]} - 9\gamma^{[2]}\widehat{G}_{\mu[2]}\right)\epsilon^* \ , \tag{3.1b}$$

$$\delta A_{\mu\nu}{}^\alpha = V_+{}^\alpha(\overline{\epsilon}^*\gamma_{\mu\nu}{}^m\lambda^*)n_m + V_-{}^\alpha(\overline{\epsilon}\gamma_{\mu\nu}{}^m\lambda)n_m$$
$$\quad - 4V_+{}^\alpha(\overline{\epsilon}\gamma_{[\mu|}{}^m\psi_{|\nu]}^*)n_m - 4V_-{}^\alpha(\overline{\epsilon}^*\gamma_{[\mu|}{}^m\psi_{|\nu]})n_m \ , \tag{3.1c}$$

$$\delta A_{\mu\nu\rho\sigma} = i(\overline{\epsilon}\gamma_{[\mu\nu\rho|}{}^m\psi_{|\sigma]})n_m - i(\overline{\epsilon}^*\gamma_{[\mu\nu\rho|}{}^m\psi_{|\sigma]}^*)n_m - \frac{3i}{8}\epsilon_{\alpha\beta} A_{[\mu\nu}{}^\alpha \delta A_{\rho\sigma]}{}^\beta \ , \tag{3.1d}$$

$$\delta\lambda = -(P_\downarrow\gamma^\mu\epsilon^*)\widehat{P}_\mu - \frac{1}{24}(P_\downarrow\gamma^{\mu\nu\rho}\epsilon)\widehat{G}_{\mu\nu\rho} \ , \tag{3.1e}$$

$$\delta V_+{}^\alpha = V_-{}^\alpha(\overline{\epsilon}^*\slashed{\eta}\,\lambda) \ , \qquad \delta V_-{}^\alpha = V_+{}^\alpha(\overline{\epsilon}\slashed{\eta}\,\lambda^*) \ , \tag{3.1f}$$

up to fermionic bilinear terms in fermionic transformation rules. Here "c.c." stands for a complex conjugation that makes the total expression real. The *hats* over field strengths and covariant derivatives denote supercovariantizations, as usual [17]. The field strengths $G_{[3]}$, *etc.* are defined in the same way as in [5][4]

$$G_{\mu\nu\rho} \equiv -\epsilon_{\alpha\beta} V_+{}^\alpha F_{\mu\nu\rho}{}^\beta \ , \qquad P_\mu \equiv -\epsilon_{\alpha\beta} V_+{}^\alpha \partial_\mu V_+{}^\beta \ , \qquad Q_\mu \equiv -i\epsilon_{\alpha\beta} V_-{}^\alpha \partial_\mu V_+{}^\beta \ ,$$
$$F_{\mu\nu\rho}{}^\alpha \equiv 3\partial_{[\mu} A_{\nu\rho]}{}^\alpha \ , \qquad F_{\mu\nu\rho\sigma\tau} \equiv 5\partial_{[\mu} A_{\nu\rho\sigma\tau]} + \frac{5i}{8}\epsilon_{\alpha\beta} A_{[\mu\nu}{}^\alpha F_{\rho\sigma\tau]}{}^\beta \ , \quad (3.2)$$

satisfying the Bianchi identities

$$D_{[\mu} P_{\nu]} = 0 \ , \qquad D_{[\mu} G_{\nu\rho\sigma]} = +P_{[\mu} G^*_{\nu\rho\sigma]} \ , \qquad \partial_{[\mu_1} F_{\mu_2 \cdots \mu_6]} \equiv \frac{5i}{12} G_{[\mu_1\mu_2\mu_3} G^*_{\mu_4\mu_5\mu_6]} \ , \quad (3.3\text{a})$$
$$\partial_{[\mu} Q_{\nu]} = -i P_{[\mu} P^*_{\nu]} \ . \quad (3.3\text{b})$$

As in the $N=1$ formulations in 12D [1][10], we need extra constraints on fields in our system, using null-vectors:

$$n^m \widehat{G}_{mnr} = 0 \ , \qquad n^m \widehat{F}_{mnrst} = 0 \ , \qquad n^m \widehat{R}_{mn}{}^{rs} = n_r \widehat{R}_{mn}{}^{rs} = 0 \ , \quad (3.4\text{a})$$
$$n^m \widehat{P}_m = 0 \ , \qquad n^m Q_m = 0 \ , \quad (3.4\text{b})$$
$$n^m \widehat{\mathcal{R}}_{mn} = 0 \ , \qquad \overline{\widehat{\mathcal{R}}}_{mn} \eta\!\!\!/ = 0 \ , \quad (3.4\text{c})$$
$$\eta\!\!\!/ \lambda = 0 \ , \qquad n^\mu \widehat{D}_\mu \lambda = 0 \ . \quad (3.4\text{d})$$

Here $\widehat{\mathcal{R}}_{\mu\nu} \equiv 2\widehat{D}_{[\mu} \psi_{\nu]}$ is the gravitino field strength. These forms are quite natural according to the experience with the $N=1$ supergravity [1] or $N=1$ supersymmetric Yang-Mills theory [10]. The constraints (3.4c) and (3.4d) are similar to those in [1]. The *bar* in (3.4c) denotes the gravitino field strength multiplied by $\eta\!\!\!/$ from the right, obeying the rule (2.5). There are also other implicit constraints related to (3.4) *via* Bianchi identities, such as $n^m \widehat{D}_m \widehat{\mathcal{R}}_{rs} = 0$, which are not written explicitly. Relevantly, our fields allow extra transformations analogous to the case of $N=1$ supersymmetric Yang-Mills in 12D [10]:

$$\delta_{\text{E}} \varphi_{\mu_1 \cdots \mu_m}{}^{m_1 \cdots m_n} = \Omega_{[\mu_1 \cdots \mu_{m-1}}{}^{m_1 \cdots m_n} n_{\mu_m]} + \Omega'_{\mu_1 \cdots \mu_m}{}^{[m_1 \cdots m_{n-1}} n^{m_n]} \ , \quad (3.5)$$

where we use the symbol $\varphi_{\mu_1 \cdots \mu_m}{}^{m_1 \cdots m_n}$ for any fundamental field in our multiplet with arbitrary number of curved indices $\mu_1, \cdots, \mu_m$ and local Lorentz indices $m_1, \cdots, m_n$.

We have checked the closure on all the bosonic fields, using also the anti-self-duality (3.7d). Relevantly the transformation rules for fermionic fields are fixed up to fermionic bilinear terms, as usual in component formulation [17] in supergravity. Even though we did not confirm the closure on fermionic fields for our peculiar system in 12D, we are confident about the consistency of the system, and there will be no disturbing problem posed by the higher-order fermionic terms, according to our experience with the $N=1$ supergravity [1].

There is a technical comment to be given related to the Lorentz generators (2.4). As was briefly mentioned in [1], even the charge conjugation matrix $C_{\alpha\beta}$ does not commute with these generators due to their peculiar property. This creates complication in confirming the commutator algebra, because of superficially "constant" matrices are no longer commuting with $\mathcal{M}_{mn}$. Fortunately, however, we can easily confirm that all the superficially "constant" matrices in our supertransformation rule (3.1), e.g., $P_\downarrow \gamma^{[5]} \gamma_m$ in $\delta\psi_\mu$ commute with $\mathcal{M}_{mn}$, up to terms that can be identified as extra transformation (3.5). To this end, useful relations are such as

$$[\omega_\mu{}^{rs}\widetilde{\mathcal{M}}_{rs}, (\gamma^{mn})_{\alpha\beta}n_n + (P_{\uparrow\downarrow})_{\alpha\beta}n^m] = 0 ,$$
$$[\omega_\mu{}^{rs}\widetilde{\mathcal{M}}_{rs}, (\not{n}\gamma^{[3]})_{\alpha\beta}]G_{[3]} = 0 , \quad [\omega_\mu{}^{rs}\widetilde{\mathcal{M}}_{rs}, (\gamma^{mn}\not{n})_\alpha{}^{\dot\beta}] = 0 , \tag{3.6}$$

---

[4]The $P_\mu$ is not confusing with $P_\uparrow$ or $P_\downarrow$, because the index $\mu$ can not be $\uparrow$ or $\downarrow$.

*etc.* in addition to (3.12) in [1]. In this manipulation, we need regard the gravitino with superscript $\psi_\mu{}^\alpha$ and gravitello with subscript $\lambda_{\dot\alpha}$ as the fundamental fields, and their lowering/raising always need $C_{\alpha\beta}$ or $C^{\alpha\beta}$.

Our field equations for the scalars, second-rank anti-symmetric tensor, zwölfbein, fifth-rank field strength, gravitino and gravitello are respectively

$$\widehat{D}_\mu \widehat{P}^\mu - \tfrac{1}{24}\widehat{G}^2_{\mu\nu\rho} + \mathcal{O}(\psi^2) = 0 , \tag{3.7a}$$

$$\widehat{D}_\mu \widehat{G}^\mu{}_{[\nu\rho}n_{\sigma]} + \widehat{P}^\mu \widehat{G}^*_{\mu[\nu\rho}n_{\sigma]} + \tfrac{2i}{3}\widehat{F}_{[\nu\rho|}{}^{\tau\omega\lambda}\widehat{G}_{\tau\omega\lambda}n_{|\sigma]} + \mathcal{O}(\psi^2) = 0 , \tag{3.7b}$$

$$\left(\widehat{R}_{\rho[\mu|} - \widehat{P}_\rho\widehat{P}^*_{[\mu|} - \widehat{P}_{[\mu|}\widehat{P}^*_\rho - \tfrac{1}{6}\widehat{F}_{[4]\rho}\widehat{F}^{[4]}{}_{[\mu|} \right.$$
$$\left. - \tfrac{1}{8}\widehat{G}_\rho{}^{\sigma\tau}\widehat{G}^*_{\sigma\tau[\mu|} - \tfrac{1}{8}\widehat{G}_{\sigma\tau[\mu|}\widehat{G}_\rho^{*\sigma\tau} + \tfrac{1}{48}g_{\rho[\mu|}\widehat{G}^{[3]}\widehat{G}^*_{[3]}\right)n_{|\nu]} + \mathcal{O}(\psi^2) = 0 , \tag{3.7c}$$

$$\widehat{F}_{[m_1\cdots m_5}n_{m_6]} = -\tfrac{1}{6!}\epsilon_{m_1\cdots m_6}{}^{p_1\cdots p_6}\widehat{F}_{p_1\cdots p_5}n_{p_6} , \tag{3.7d}$$

$$\not{n}\left(\gamma^\rho\widehat{\mathcal{R}}_{\rho[\mu|} - \lambda^*\widehat{P}_{[\mu|} - \tfrac{1}{48}\gamma^{[3]}\gamma_{[\mu|}\lambda\widehat{G}^*_{[3]} - \tfrac{1}{96}\gamma_{[\mu|}\gamma^{[3]}\lambda\widehat{G}^*_{[3]}\right)n_{|\nu]} = 0 , \tag{3.7e}$$

$$\not{n}\left(\gamma^\mu\widehat{D}_\mu\lambda - \tfrac{i}{240}\gamma^{[5]}\lambda\widehat{F}_{[5]}\right) = 0 , \tag{3.7f}$$

up to fermionic bilinear terms $\mathcal{O}(\psi^2)$ in the bosonic field equations.

First brief remarks are in order about the derivations of these field equations. The procedure of getting these field equations is more or less parallel to the 10D case [5], namely we first postulate the gravitello and gravitino field equations with the structures in (3.7f) and (3.7e) with unknown coefficients, and then take their supersymmetry variations under (3.1). One of the crucial ansätze made is the anti-self-duality condition (3.7d) for the fifth-rank antisymmetric field strength, which is now formally "oxidized" in 12D to $\mathcal{F}_{[6]} = -\epsilon_{[6]}{}^{[6]'}\mathcal{F}_{[6]'}$ for $\mathcal{F}_{m_1\cdots m_6} \equiv F_{[m_1\cdots m_5}n_{m_6]}$.[5] We have found that this form is the consistent postulate to be used at various steps in the derivations of other bosonic field equations out of the supersymmetric variations of fermionic field equations. As for the possible structures of putting the null-vectors, we took trial and error process, until we get the result that all the unwanted terms causing troubles in the dimensional reductions into 10D (as will be performed shortly) are cancelled by themselves. Another guiding principle is to rely on the $N = 1$ results [1], which gives us the basic structures of the system, such as the zwölfbein transformation rule with the peculiar involvement of null-vector.

---

[5]Due to the choice of our signature in 12D, the self-duality condition in 10D [5] is oxidized to anti-self-duality in 12D.

There are some similarities of this result to the $N=1$ supergravity in 12D [1], as well as differences. The second-rank anti-symmetric tensor (3.7b), the zwölfbein (3.7c), gravitino (3.7d) and gravitello (3.7f) field equations look much like those in the $N=1$ case [1][10], while the scalar field equation (3.7a) and the anti-self-duality condition (3.7d) in our $N=2$ system are new in 12D. Similarities of this system to that of type IIB chiral supergravity in 10D [5] are such as the lack of invariant lagrangian, or the scalar field equation (3.7a) exactly in the same form as in 10D case [5]. One important difference is the structure of the anti-self-duality condition (3.7d) which now has six free indices with the null-vector involved.

We now give detailed technical remarks for the derivation of our result (3.1) and (3.7). Since the present formulation is the component formulation, it is easier to fix first the transformation rule (3.1) at the lowest order. As has been already mentioned for the closure check, we postulated that the fourth-rank tensor field should satisfy the anti-self-duality (3.7d) which is easily seen to be re-produce the self-duality in 10D [5]. We next fixed the linear-order transformation rule in (3.1) by demanding the closure of two supersymmetries, relying on the $N=1$ result $T_{\alpha\beta}{}^c = (\gamma^{cd})_{\alpha\beta} n_d + (P_{\uparrow\downarrow})_{\alpha\beta} n^c$ [1]. Effectively the last term here is not essential due to the extra symmetry (3.5) for the zwölfbein. The closure of two supersymmetries should be realized up to the extra transformations (3.5) for each field. By this requirement, together with the known-result for the 10D case [5] that is supposed to be coming out by simple dimensional reduction, we fixed all the structures of these transformations. In particular, the complex conjugation rules or flipping rules (2.13) and (2.14) in 12D play crucial roles in this process.

Once the transformation rule is fixed at the linear-order, we can next derive the field equations. We first postulate the gravitello field equation with the structure (3.7f) with unknown coefficients, like in [5]. These constants are fixed by the requirement that the supersymmetric variation of the gravitello field equation produce the scalar $V_\pm{}^\alpha$ and the tensor $A_{\mu\nu}{}^\alpha$-field equations. We found that the presence of $\slashed{n}$ in front is essential to delete undesirable terms left over in these bosonic field equations after the supersymmetric variations, which do not reproduce the 10D result [5] by dimensional reduction. As in [5], the supersymmetric variation of (3.7f) yields two sectors, the $\epsilon$-dependent sector and the $\epsilon^*$-dependent sector, where $\epsilon$ and $\epsilon^*$ are the parameters of supersymmetry transformation. The former yields the $A_{\mu\nu}{}^\alpha$-field equation, while the latter the $V_\pm{}^\alpha$-field equation. In this process, the coefficient of the $\lambda F$-term is consistently fixed by these two sectors. This process is just 12D analog to the similar process in [5] such as usage of the anti-self-duality (3.7d), together with $\gamma$-matrix algebra in 12D, *etc*. Some peculiar relations for the null-vectors are also used here: $[P_\uparrow, \gamma_r] = -[P_\downarrow, \gamma_r] = m_r \slashed{n} - n_r \slashed{m}$.

The gravitino field equation is postulated to have the structure as in (3.7e) with four unknown coefficients like $b_1, \cdots, b_4$ in [5], that in turn produces zwölfbein field equation under supersymmetry. As in the gravitello case, the supersymmetric variation produces both $\epsilon^*$ and $\epsilon$-linear terms. The $\epsilon^*$-terms generate $DG$, $PG^*$ and $FG$-terms, where some parts of the $DG$-terms communicate with the $PG^*$-terms due to the Bianchi identity (3.2).

The $DG$ and $PG^*$-terms fix all the four coefficients consistently, while all the $FG$-terms cancel themselves. One important point here is that the $\gamma^{[7]}FG$ and $\gamma^{[3]}FG$-terms talk to each other under the anti-self-duality property of the field strength $\mathcal{F}_{\mu_1\cdots\mu_6} \equiv F_{[\mu_1\cdots\mu_5}n_{\mu_6]}$. What happens here is just similar to the 10D case [5] despite of the 12D indices, due to the presence of $\not{n}$ or $P_\downarrow$, as a common factor stuck in front or the middle of gamma-matrices. For example, we have $\not{n}\gamma^\rho P_\downarrow \gamma_\rho = 10\not{n}$, instead of $12\not{n}$. This is the important feature of the null-vectors, that enables us to obtain a parallel structure to the 10D case, in particular the coefficients in the field equations, in spite of 12D indices. In this $\epsilon^*$-sector, we have also to use the $A_{\mu\nu}{}^\alpha$-field equation already obtained from the gravitello field equation, which constitutes a convenient cross-check. Finally the $\epsilon$-sector produces the zwölfbein field equation (3.7c), after yielding the conditions on the four coefficients, which are consistent with the $\epsilon^*$-sector. The total $\epsilon$-sector consists of Ricci-tensor terms, $PP^*$-terms, $\partial F$-terms, $FF$-terms, and $GG^*$-terms. For the $FF^*$-term, we use various duality relations to have dramatic cancellations among $\gamma^{[7]}\epsilon FF$, $\gamma^{[3]}\epsilon FF$ and $\gamma^{[5]}\epsilon FF$-terms, which turns out to be just parallel to the 10D case, leaving only $\gamma^{[1]}\epsilon FF$-terms contributing to the zwölfbein field equation. Similarly, $\partial F$- terms and $GG^*$-terms talk to each other through the Bianchi identity (3.2), and after the similar cancellations among $\gamma^{[7]}\epsilon GG^*$, $\gamma^{[5]}\epsilon GG^*$, $\gamma^{[3]}\epsilon GG^*$, we see that only the $\gamma^{[1]}\epsilon GG^*$-terms contribute to the zwölfbein field equations as desired, and also similarly to the 10D case [5].

Even though all the details of these processes seem rather complicated, once we understand the parallel feature of our system to the 10D case, we easily recognize the similarity of our 12D theory to the type IIB theory in 10D [5].

### 4. Dimensional Reduction into Type IIB Supergravity in 10D

Our next crucial check is whether our system reproduces the known results in lower dimensions upon appropriate dimensional reductions. Here we perform a dimensional reduction into 10D, in order to reproduce the well-known results in type IIB theory [5]. This process is, however, already partially performed, when we had fixed the postulate for the fermionic field equations with technical points in the previous section. So we give here only crucial ingredients briefly.

The important setup is the notational conventions, such as indices to distinguish the 10D ones from 12D ones. We follow the same notation as in ref. [1], namely we use only in this section the *hats* for fields and indices in 12D, while those in 10D have no *hats*. Other prescriptions for the dimensional reduction are the usual ones, such as the independence of all the fields on the extra dimensions, such as $\widehat{G}_{\mu\nu\pm} = 0$, $\widehat{D}_\pm = 0$. Relevant and frequently used equations are such as [10]

$$\widehat{\gamma}_{\widehat{m}} = \begin{cases} \widehat{\gamma}_m = \gamma_m \otimes \sigma_3 \ , \\ \widehat{\gamma}_{(11)} = I \otimes \sigma_1 \ , \\ \widehat{\gamma}_{(12)} = -I \otimes i\sigma_2 \ , \end{cases} \quad \widehat{\not{n}} = I \otimes \begin{pmatrix} 0 & 1 \\ 0 & 0 \end{pmatrix} \ , \quad \widehat{\not{m}} = I \otimes \begin{pmatrix} 0 & 0 \\ 1 & 0 \end{pmatrix} \ ,$$

$$\widehat{C} = -\widehat{\gamma}_{(0)}\widehat{\gamma}_{(12)} = \gamma_{(0)} \otimes \sigma_1 = C \otimes \sigma_1 \quad , \quad \widehat{\gamma}_{13} \equiv \widehat{\gamma}_{(0)}\widehat{\gamma}_{(1)} \cdots \widehat{\gamma}_{(9)}\widehat{\gamma}_{(11)}\widehat{\gamma}_{(12)} = \gamma_{11} \otimes \sigma_3 \quad ,$$

$$\widehat{P}_{\uparrow} = I \otimes \begin{pmatrix} 1 & 0 \\ 0 & 0 \end{pmatrix} \quad , \quad \widehat{P}_{\downarrow} = I \otimes \begin{pmatrix} 0 & 0 \\ 0 & 1 \end{pmatrix} \quad , \quad \overline{\widehat{\mathcal{R}}}_{\widehat{\mu}\widehat{\nu}} = (\overline{\mathcal{R}}_{\widehat{\mu}\widehat{\nu}}, 0) \quad , \quad \widehat{\lambda} = \begin{pmatrix} 0 \\ \lambda \end{pmatrix} \quad , \quad (4.1)$$

where $\sigma_1$, $\sigma_2$, $\sigma_3$ are the standard $2 \times 2$ Pauli matrices. Other related notations in 10D are such as $(\eta_{mn}) = \text{diag.}(-,+,\cdots,+)$, $\epsilon^{01\cdots 9} = +1$, $\gamma_{11} = \gamma_{(0)}\gamma_{(1)}\cdots\gamma_{(9)}$. Note that the chiralities of the spinors in 10D will be $\gamma_{11}(\psi_\mu, \lambda) = (\psi_\mu, -\lambda)$.

Our goal is now to re-derive the field equations for the type IIB theory [5]. We start with the scalar field eq. (3.7a) which directly yields the same scalar field equation in 10D [5] up to notational differences such as the space-time signature, and there is no other indices to be considered. The next one is the second-rank anti-symmetric tensor field equation (3.7b) which gives eq. (5.2) in [5] for the indices $\widehat{\mu} = \mu$, $\widehat{\nu} = \nu$, $\widehat{\rho} = \rho$, $\widehat{\sigma} = +$. Other index combinations, such as $\widehat{\mu} = \mu$, $\widehat{\nu} = \nu$, $\widehat{\rho} = +$, $\widehat{\sigma} = +$ give only vanishing results, under our dimensional reduction prescription above as well as our constraints (3.4). The zwölfbein field eq. (3.7c) yields the zehnbein field equation (5.3) in [5] for the index combination $\widehat{\rho} = \rho$, $\widehat{\mu} = \mu$, $\widehat{\nu} = +$, while other indices give vanishing results. Interestingly, the anti-self-duality condition (3.7d) yields the desirable self-duality condition (5.4) in ref. [5] for our indices $\widehat{\mu}_1 = \mu_1$, $\cdots$, $\widehat{\mu}_5 = \mu_5$, $\widehat{\mu}_6 = +$, while other indices give trivial results $0 = 0$. As for the gravitino field eq. (3.7e), we can get the 10D gravitino field eq. (4.12) in ref. [5] for the indices $\widehat{\mu} = \mu$, $\widehat{\nu} = +$, and other components give trivial equations. Lastly, the gravitello field eq. (3.7f) yields the corresponding eq. (4.6) in [5], and there is no other component to be considered.

## 5. $N = 2$ Supergravity in 12D for Super $(2+2)$-Brane

Once we have established component formulation of $N = 2$ supergravity in 12D, we can easily re-formulate the same system in superspace, and we can further formulate super $(2+2)$-brane [7] on such 12D background, as we did similar analysis for Green-Schwarz superstring for $N = 1$ supergravity in 12D [1].

Among superspace constraints, the lower-dimensional superspace constraints with the dimensionality $d = 0$ are easy to extract out of our transformation rule (3.1), following the general technique in [18]. In this paper we give only the $d = 0$ constraints relevant to the fermionic invariance of super $(2+2)$-brane action:

$$T_{\alpha\beta}{}^c = (\gamma^{cd})_{\alpha\beta} n_d + (P_{\uparrow\downarrow})_{\alpha\beta} n^c \quad , \quad T_{\overline{\alpha}\overline{\beta}}{}^c = (\gamma^{cd})_{\alpha\beta} n_d + (P_{\uparrow\downarrow})_{\alpha\beta} n^c \quad , \quad (5.1a)$$

$$F_{\alpha\beta cde} = -\frac{i}{4}(\gamma_{cde}{}^f)_{\alpha\beta} n_f \quad , \quad F_{\overline{\alpha}\overline{\beta} cde} = +\frac{i}{4}(\gamma_{cde}{}^f)_{\alpha\beta} n_f \quad . \quad (5.1b)$$

We use only in this section the superspace notation as in refs. [1][10], such as $A, B, \cdots = (a, \alpha, \overset{\bullet}{\alpha}, \overline{\alpha}, \overline{\overset{\bullet}{\alpha}})$, $(b, \beta, \overset{\bullet}{\beta}, \overline{\beta}, \overline{\overset{\bullet}{\beta}})$, $\cdots$ with $a, b, \cdots = (0), (1), \cdots, (9), (11), (12)$ for the local bosonic indices, while $\alpha, \beta, \cdots = 1, 2, \cdots, 32$ (*idem. for* $\overline{\alpha}, \overset{\bullet}{\alpha}, \overline{\overset{\bullet}{\alpha}}$) for the chiral spinorial indices. Since the $\gamma$-matrices "do not know" the *barred-ness* of the spinors to be multiplied like the 10D case [19], it is not necessary to put the *bars* on the spinorial indices of the $\gamma$-matrices in (5.1).[6]

Our total action $S$ for super $(2+2)$-brane is similar to that for usual $p$-brane [12], or it is a curved superspace generalization of that in [7]:

$$S = S_\sigma + S_A \ , \tag{5.2a}$$

$$S_\sigma \equiv \int d^4\sigma \left( \tfrac{1}{2} \sqrt{g} g^{ij} \eta_{ab} \Pi_i{}^a \Pi_j{}^b - \sqrt{g} \right) \ , \tag{5.2b}$$

$$S_A \equiv \int d^4\sigma \left( -\tfrac{i}{6} \epsilon^{i_1 \cdots i_4} \Pi_{i_1}{}^{B_1} \cdots \Pi_{i_4}{}^{B_4} A_{B_4 \cdots B_1} \right) \ , \tag{5.2c}$$

Here $g \equiv \det(g_{ij})$ is the determinant of the metric $g_{ij}$ on the $(2+2)$-dimensional world-supervolume, $\eta_{ab}$ is the bosonic component of the 12D superspace metric, and $\Pi_i{}^A \equiv (\partial_i Z^M) E_M{}^A$ is usual pull-back used in the $p$-brane formulation [12]. Since we have the $(2,2)$ signature, we have the positive definite signature $g = \det(g_{ij}) > 0$. The $i, j, \cdots = 0, 1, 2, 3$ are the curved indices on the $(2,2)$-dimensional world-supervolume, while $(i), (j), \cdots = (0), (1), (2), (3)$ are for the local Lorentz indices, with the flat metric $(\eta_{(i)(j)}) = \text{diag.}(-,+,+,-)$. Compared with the general $p$-brane formulation [12], the new feature here is the signature $(2,2)$ for the world-supervolume reflected in the presence of the imaginary unit $i$ in (5.2c), induced by a Wick rotation from the $(3,1)$-signature.

We now postulate our local fermionic $\kappa$ [11] and $\eta$-transformations for our total action:

$$\delta E^\alpha = (I+\Gamma)^\alpha{}_\beta \kappa^\beta + \tfrac{1}{2} (\not{\eta}\not{\eta})^{\alpha\beta} \eta_\beta \equiv -[(I+\Gamma)\kappa]^\alpha + (P_\uparrow \eta)^\alpha \ , \tag{5.3a}$$

$$\delta \overline{E}^{\overline{\alpha}} = (I+\Gamma)^\alpha{}_\beta \overline{\kappa}^{\overline{\beta}} + \tfrac{1}{2} (\not{\eta}\not{\eta})^{\alpha\beta} \overline{\eta}_{\overline{\beta}} \equiv -[(I+\Gamma)\overline{\kappa}]^{\overline{\alpha}} + (P_\uparrow \overline{\eta})^{\overline{\alpha}} \ , \quad \delta E^a = 0 \ , \tag{5.3b}$$

where as usual $\delta E^A \equiv (\delta Z^M) E_M{}^A$, and

$$\Gamma \equiv \tfrac{1}{24\sqrt{g}} \epsilon^{ijkl} \Pi_i{}^a \Pi_j{}^b \Pi_k{}^c \Pi_l{}^d (\gamma_{abcd}) \ . \tag{5.4}$$

We do not have to vary the metric $g_{ij}$, when confirming the invariance of the action, due to its algebraic field equation for the usual embedding condition $g_{ij} = \eta_{ab} \Pi_i{}^a \Pi_j{}^b$ [12]. As in the general $p$-brane formulation [12], we can confirm the relations such as

$$\Gamma^2 = I \ ,$$

$$\tfrac{1}{\sqrt{g}} \epsilon_i{}^{jkl} \Pi_j{}^a \Pi_k{}^b \Pi_l{}^c \gamma_{abc} \Gamma = +6 \Pi_i{}^a \gamma_a \ ,$$

$$\Pi_i{}^\alpha \Pi^{ib} (\gamma_b \Gamma) = \tfrac{1}{6\sqrt{g}} \epsilon^{ijkl} \Pi_i{}^\alpha \Pi_j{}^b \Pi_k{}^c \Pi_l{}^d (\gamma_{bcd}) \ . \tag{5.5}$$

As usual, the first and last ones follow the second one, which is the most fundamental.

---

[6] The *bars* on the spinorial indices in 12D should not be confused with the *dots* which denote the chirality of the spinors [10]. This is similar to the type IIB supergravity in 10D [19].

The invariance check of our total action $S$ under (5.3) goes in a way similar to the general $p$-brance case [12], because for general variations $\delta E^\alpha \neq 0$, $\delta \overline{E}^{\overline{\alpha}} \neq 0$, $\delta E^a = 0$ we have

$$\delta\left(S_\sigma + S_A\right) = \left[\sqrt{g}g^{ij}\Pi_i{}^\gamma \left(\gamma_a \not{n}\right)_{\gamma\beta}\left(\delta E^\beta\right) \Pi_j{}^a \right.$$
$$\left. + \frac{1}{6}\epsilon^{ijkl}\Pi_i{}^\gamma \left(\gamma_{bcd}\not{n}\right)_{\gamma\alpha}\left(\delta E^\alpha\right)\Pi_j{}^b\Pi_k{}^c\Pi_l{}^d\right] + (\delta E^\alpha \to \delta\overline{E}^{\overline{\alpha}}) \ . \tag{5.6}$$

Now it is clear that the total action is invariant under our $\eta$-transformation in (5.3) due to $\not{n}P_\uparrow \equiv 0$, while for the $\kappa$-transformation [11][12] we need (5.5) as well as an extra constraint like the $N=1$ case [1],

$$\Pi_i{}^a n_a = 0 \ , \tag{5.7}$$

which gets rid of all the unwanted terms. This constraint is formally the same as that arose from the constraint lagrangian in the $N=1$ case in 12D [1]. Since our world-supervolume is $(2+2)$-dimensional, we can impose this constraint from outside, *unlike* the case of Green-Schwarz superstring [1] with 2D world-sheet, where such a constraint is regarded as a field equation not to be imposed by hand, when confirming action invariance. As in ref. [1], we can also confirm the invariance of the constraint (5.7) itself under (5.3) due to $T_{\alpha b}{}^d = 0$.

As in the $N=1$ case [1], the role of these two fermionic symmetries is to delete the unwanted degrees of freedom in the original 32-components in 12D for each Majorana-Weyl spinors, schematically expressed as $32 \xrightarrow{\eta} 16 \xrightarrow{\kappa} 8$, where the last $\kappa$-symmetry is playing exactly the same role as in type IIB superstring formulation [11], while the $\eta$-symmetry [1] is getting rid of the unwanted doubling of the degrees of freedom in 12D compared with 10D.

The action $S_A$ with the four indices arising from the 4D world-supervolume is the natural goal for the super $(2+2)$-brane on a general curved superspace background with the fourth-rank potential $A_{[4]}$. To put it differently, the presence of the fourth-rank antisymmetric tensor in type IIB supergravity in 10D indicates the existence of consistent formulation of 3-brane or super $(2+2)$-brane formulation with 4D world-supervolume *via* Wess-Zumino-Novikov-Witten term, as the supermembrane theory [20] was indicated by the third-rank tensor in supergravity in 11D.

## 6. Concluding Remarks

In this paper we have established an $N=2$ chiral supergravity in 12D for the first time. Our result has similarity to as well as difference from the $N=1$ case [1] or from the type IIB supergravity in 10D [5]. We have seen interesting involvement of the null-vector $n_\mu$ in the anti-self-duality condition for the fifth-rank antisymmetric field strength, which reproduces the familiar self-duality condition in 10D [5] upon simple dimensional reduction. This result is very similar to the $N=1$ case [1][10], where the parallel structure arose in the process. This also provides a strong confirmation for the consistency of our system.

We have also shown how our $N=2$ chiral supergravity can be the consistent superspace backgrounds for super $(2+2)$-brane [7]. We have seen that under a particular constraint $\Pi_i{}^a n_a = 0$, our total action is invariant under the fermionic $\eta$ and $\kappa$-symmetries. This result gives the direct confirmation of the validity of our system as the weak coupling limit of F-theory [2] with the right degrees of freedom. Our present result together with our previous one for the $N=1$ case [1] justify our elaborate usage of the null-vectors in these supergravity formulations.

One subtlety we encountered in our formulation is about the coset $SU(1,1)/U(1)$, which is built-in from the outset. From the F-theory viewpoint [2][16], however, this coset most likely comes out of compactifications, such as toroidal compactifications. There may well be some freedom in formulating $N=2$ supergravity in 12D with null-vectors, that we do not know yet. This question is still to be answered in future studies.

Even though we did not try in this paper, we can further try the following studies: We can truncate our $N=2$ theory into $N=1$ theory [1] within 12D, by deleting halves of the gravitini and gravitelli, the fifth-rank tensor, and one of the scalar fields. We can also perform "double-dimensional reduction" [21] of super $(2+2)$-brane in 12D into superstring in 10D. We can also complete the superspace formulation in our 12D by extracting superspace constraints using the general method in [18]. We can also consider compactifications of our 12D theory on elliptic Calabi-Yau manifold with hodge numbers $h^{1,1}$ and $h^{2,1}$ to get $N=1$ supergravity in 6D [2]. With $V$, $T$ and $H$ respectively for the numbers of vector, tensor and hypermultiplets in 6D supergravity, we will get $V+T = h^{1,1} - 2$ and $H = h^{2,1} + 1$, which can be described by the most general couplings explored recently in ref. [22]. Additionally we can explore the relationship between the $(2+2)$-dimensional world-supervolume and (supersymmetric) self-dual theories which is the target space-time for $N=2$ superstring [2][23]. It may well be that even F-theory itself [2] is a different manifestation of target space-time physics in $N=2$ superstring [3].

We believe that our result is the first important step for understanding the whole structure of F-theory, which is supposed to be the ultimate master theory for other superstring theories such as type IIB or heterotic superstring in 10D. Even though our result is only for the weak coupling limit, we can not stress enough the usefulness of this result for further applications, such as investigations of compactifications [13] into lower dimensions.

Emphasis on our $N=2$ supergravity in 12D as the master theory of other superstring/supergravity theories does *not* necessarily exclude other possible higher-dimensional supergravities, *e.g.*, in 14D. There has been already some symptom of such theories, *e.g.*, by supersymmetric Yang-Mills theory [24], which suggests the existence of supergravity in $(11+3)$-dimensions. It is also interesting to see if there is a "maximal" supergravity in higher dimensions, when we allow the usage of null-vectors, like the 11D supergravity when no null-vector is allowed. Studies in these directions are now under way [25].

We are grateful to S.J. Gates, Jr., E. Sezgin, and C. Vafa for important discussions.